\documentclass{caosp309}
\pdfoutput=1
\usepackage{graphicx}
\usepackage{natbib}
\usepackage[bottom=1in]{geometry}
\usepackage[utf8]{inputenc}
\usepackage[T1]{fontenc}
\pubyear{2025}
\volume{55}
\volnumber{3}
\firstpage{357}
\received{November 14, 2024}
\accepted{December 20, 2024}

\def\BibTeX{{\rm B\kern-.05em{\sc i\kern-.025em b}\kern-.08em
             T\kern-.1667em\lower.7ex\hbox{E}\kern-.125emX}}

\begin{document}

\hauthor{B.\,Keskin and \"O.\,Ba\c{s}t\"urk}

\title{Eclipsing binary classification with machine learning techniques}

\author{B.\,Keskin\inst{1}
      \and
        \"O.\,Ba\c{s}t\"urk\inst{2,3}
       }

\institute{
           Ankara University, Graduate School of Natural \& Applied Sciences, Astronomy \& Space Sciences Department, Ankara, T\"urkiye
           \and
           Ankara University, Faculty of Science, Astronomy \& Space Sciences Department,\\ Ankara, T\"urkiye
           \and
           Ankara University, Astronomy \& Space Sciences Research \& Application Center,\\
           Kreiken Observatory, Ankara, T\"urkiye
          }


\maketitle

\begin{abstract}
We focus on the automated classification of eclipsing binary stars using deep learning methods to handle the vast data generated by large-scale photometric sky surveys. These surveys produce extensive datasets that are impractical for manual analysis. By using machine learning to classify eclipsing binary stars based on light curve morphology, this study aims to contribute to the efforts to efficiently process and accurately interpret massive data from projects Kepler, TESS and Gaia missions.
\end{abstract}

\section{Introduction}
\label{intr}
Eclipsing binaries, whose light curves show brightness variations from mutual eclipses, have large sets of photometric survey data. Automated data processing, leveraging supervised and unsupervised machine learning (ML) are essential to efficiently analyze these massive datasets and identifying patterns in time-series data.

\citet{Daza_2023} utilized ML to classify eclipsing binary stars (EBs) in the VISTA Variables of the Vía L\'{a}ctea Survey (VVV), revealing time-series features in light curves and introducing a Compound Decision Tree (CDT) model for their classification. \citet{Cokina_2021} classified eclipsing binary light curves into detached and over-contact classes. A hybrid of Bidirectional Long Short Term Memory (BiLSTM) and one dimensional Convolutional Neural Network (1D CNN), achieved 98\% accuracy, and reached 100\% when semi-detached binaries were excluded. \citet{Bodi_2021} used the Locally Linear Embedding (LLE) algorithm for classifying the Optical Gravitational Lensing Experiment (OGLE) eclipsing binary light curves based on their morphology. \citet{Suveges_2017} applied ML methods, including Functional Principal Component Analysis (FPCA), Linear Discriminant Analysis (LDA), Random Forest (RF), and Self-Organizing Map (SOM), to classify eclipsing binaries based on light curve morphology using datasets from Catalog and Atlas of Eclipsing Binaries (CALEB), High Precision Parallax Collecting Satellite (HIPPARCOS), and Kepler. \citet{Kochoska_2017} proposed a combination of the t-distributed Stochastic Neighbor Embedding (t-SNE) and Density-Based Spatial Clustering of Applications with Noise (DBSCAN) algorithms for the purposes of eclipsing binary light curve classification. The polynomial chain (polyfit) and two-Gaussian models are used to characterize the geometry of the folded light curves. Classification is done according to the morphology parameter for a given system.

This study aims to contribute to the efforts in reliable and rapid classification of eclipsing binaries, enabling statistically reproducible results across vast astronomical databases. We classify a limited sample of eclipsing binary star candidates in the Gaia DR3 archive. We utilize Transiting Exoplanet Survey Satellite (TESS)\footnote{\url{https://tessebs.villanova.edu}} and Kepler\footnote{\url{https://keplerebs.villanova.edu}} light curve data in the Villanova eclipsing binary catalogs, where the systems are ready to be labeled according to the morphology parameter \citep{Matijevic_2012}, to train our CNN model. We then apply the trained model to classify our sample of eclipsing binaries from the Gaia Data Release 3 (DR3) archive.

\section{Data and Methods}
Villanova Kepler and TESS eclipsing binary archives are used for training ML model. 2907 Kepler and 4349 TESS EB light curves are taken and labeled according to morphology parameter by the following criteria \citep{Matijevic_2012};
\vspace{1mm}

morph $< 0.5$ detached (D)

$0.5 <$ morph $< 0.7$ semidetached (SD)

$0.7 <$ morph $< 0.8$ overcontact (OC)

$0.8 <$ morp ellipsoidal (E)
\vspace{3mm}

\begin{figure}
\centerline{\includegraphics[width=0.45\textwidth,clip=]{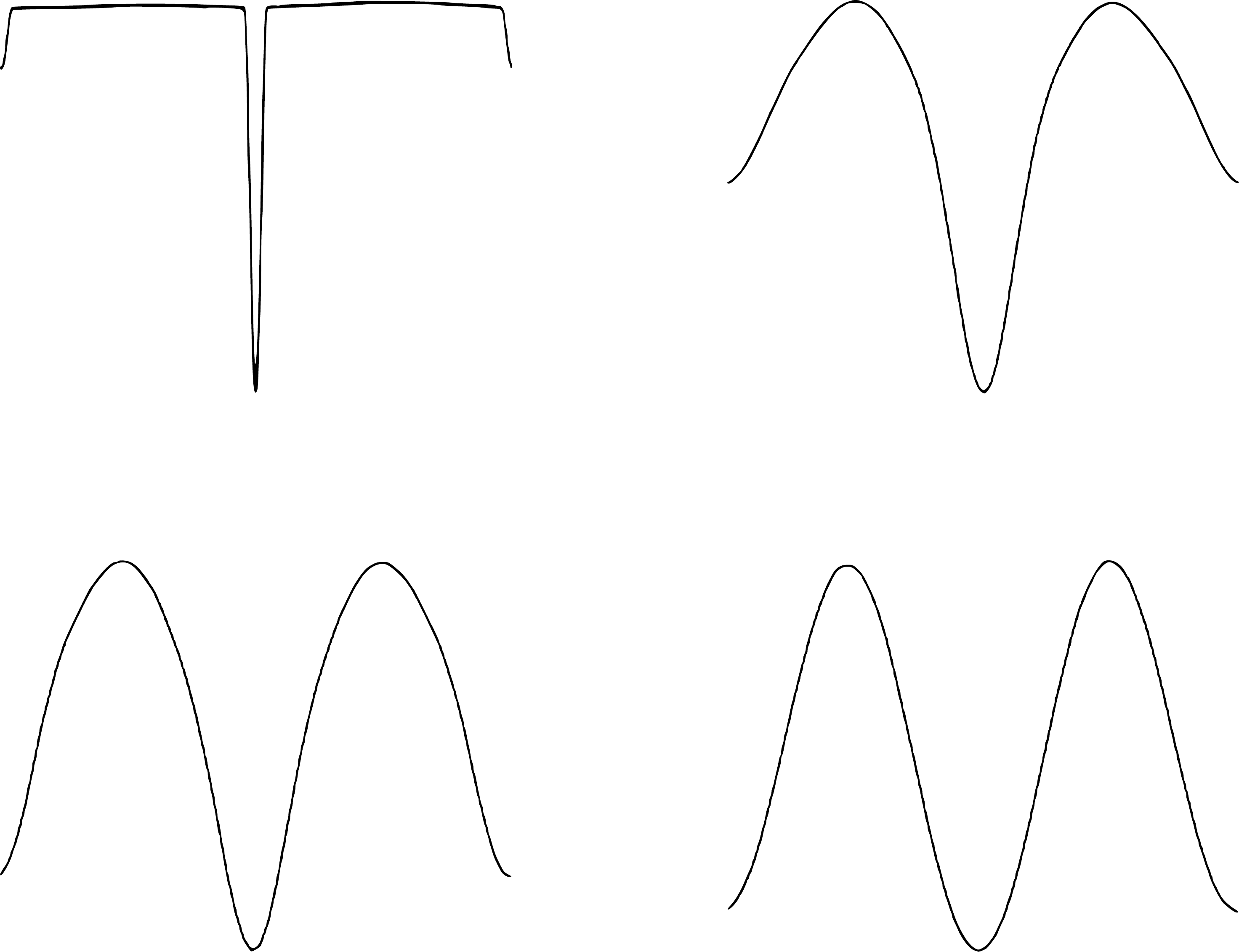}}
\caption{Sample detached (top left), semidetached (top right), overcontact (bottom left), ellipsoidal (bottom right) Kepler and TESS light curves.}
\label{sampleKeplerAndTess}
\end{figure}

The light curves are phase-folded by using light elements taken from Villanova archives and binned uniformly to standardize their resolution, ensuring consistency across datasets (Fig.\ref{sampleKeplerAndTess}).

\begin{figure}[!h]
\centering
\includegraphics [width=0.45\textwidth]{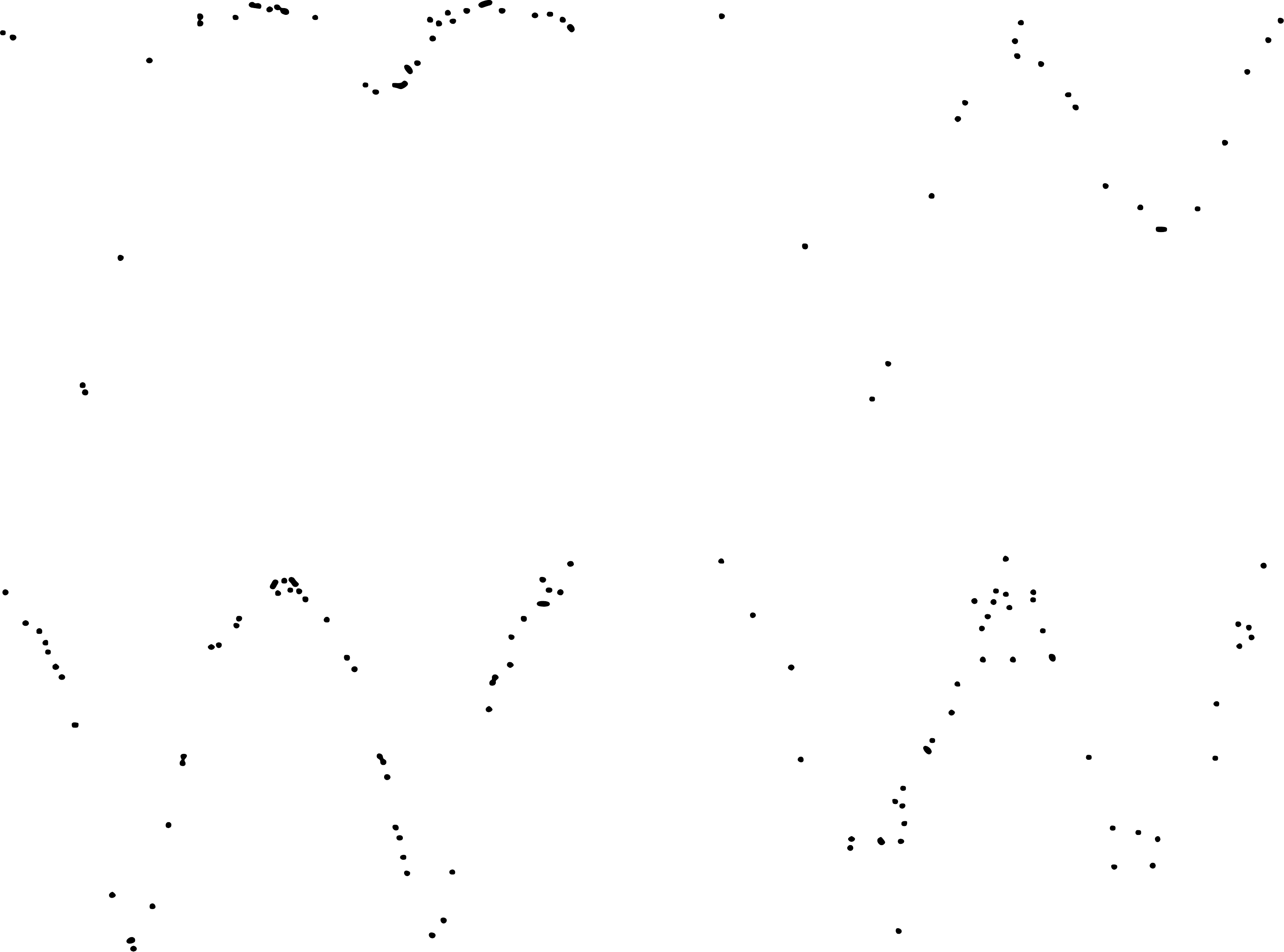}\hfill
\includegraphics [width=0.45\textwidth] {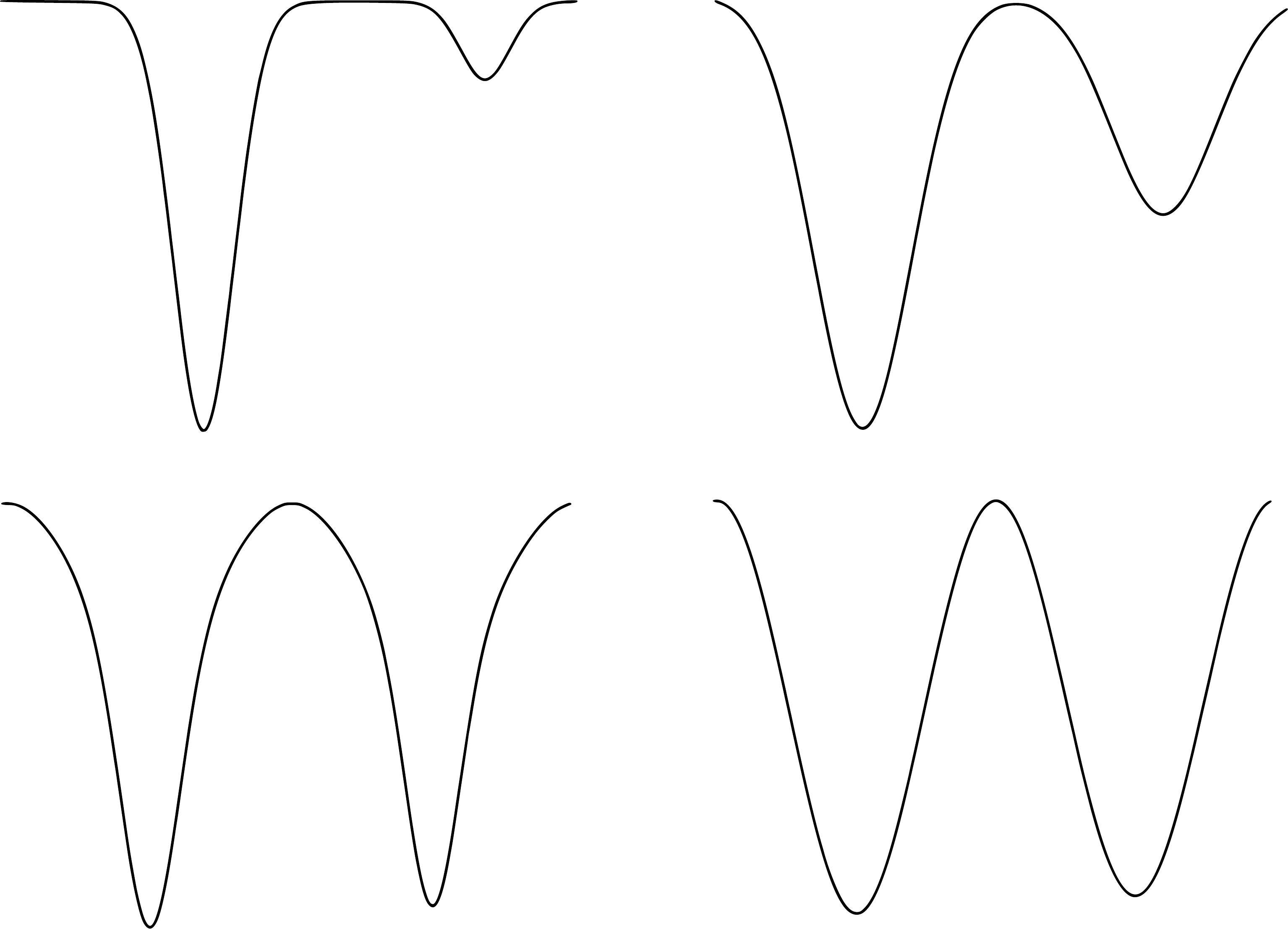}
\caption{Sample Gaia DR3 light curves (left) and their modeled light curves (right).}
\label{GaiaDR3LCsAndModeled}
\end{figure}

For classification, a small sample including 2106 EB’s were selected from Gaia DR3 archive. We provide a few of these binned and phase-folded light curves in Fig.\ref{GaiaDR3LCsAndModeled} (left) as examples. Light curves had to be processed based on two Gaussian and a cosine function due to their sampling and intensity scaling as presented in  \citet{Mowlavi_2023} (Fig.\ref{GaiaDR3LCsAndModeled} right).

The selected binaries are also in the Kepler or TESS archive. Out of 2106 EB’s 789 are from Kepler archive and 1317 are from TESS archive. So we can crossmatch the labeled and the predicted classes.

We converted light curve data of Kepler and TESS EBs to Portable Network Graphic (PNG) image files by a Python code we developed for the task. Then these light curve images were splitted into 3 groups: 67\% for training, 25\% for validation and 8\% for test. We used VGG-19 as the ML model. VGG-19 is a pretrained CNN from Visual Geometry Group (VGG) Department of Engineering Science, Oxford University. The number 19 stands for the number of layers with trainable weights. We chose VGG-19 for its ability to effectively extract nuanced features from light curve data, such as subtle variations in amplitude, shape and periodicity. VGG-19’s fine-tuning, combined with the use of regularization techniques, provided results without overfitting. We used "reduce LR on plateau" and "early stopping" methods to avoid overfitting. We also conducted experiments with varying hyperparameters and confirmed that overtraining did not occur.

\section{Results and Discussion}
\begin{figure}
\centerline{\includegraphics[width=0.45\textwidth,clip=]{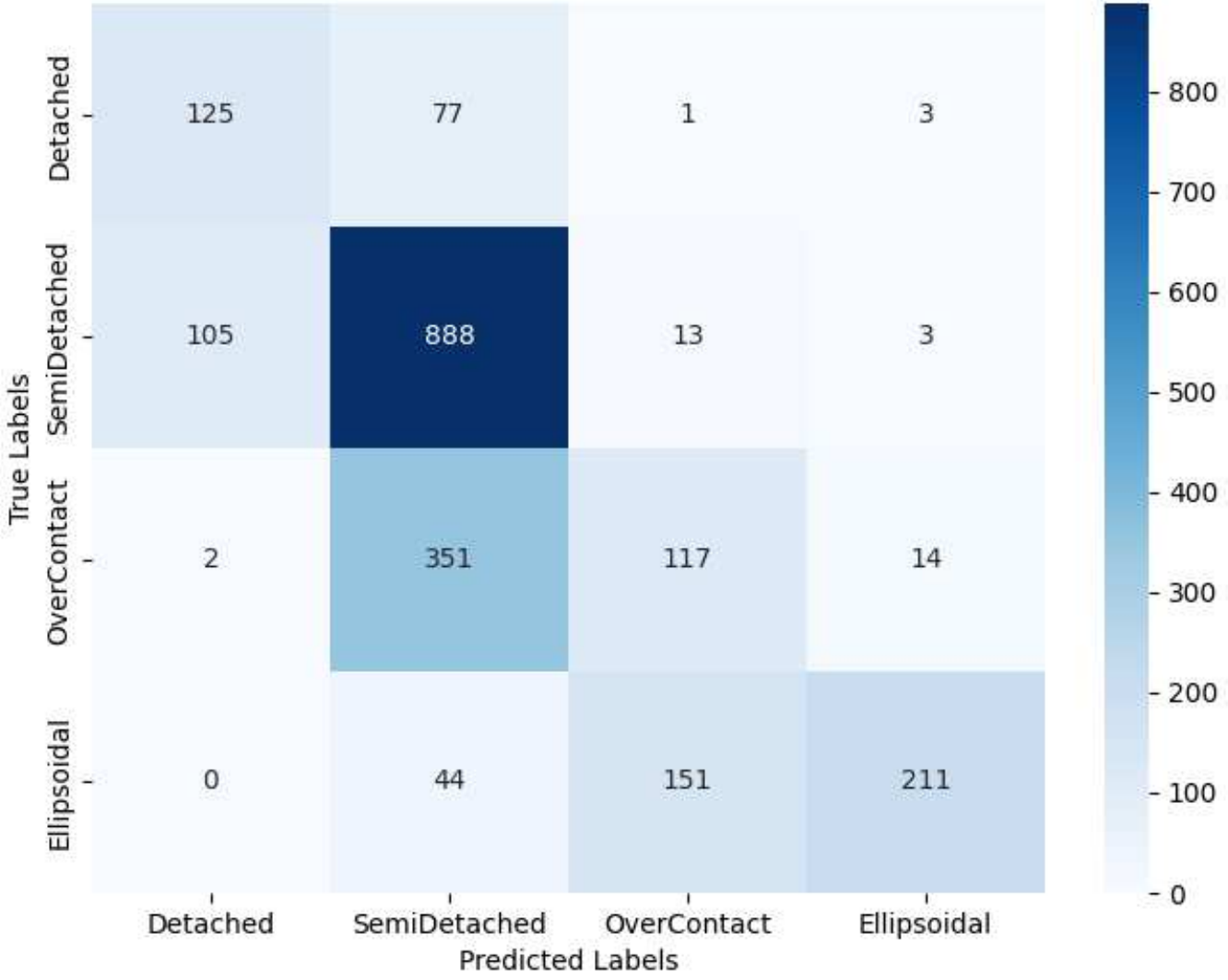}}
\caption{Confusion matrix.}
\label{confmatrix}
\end{figure}

We achieved 91\% accuracy on Kepler and TESS test data and 64\% accuracy on Gaia DR3 data as given with the confusion matrix in Fig.\ref{confmatrix}. Our ML model is highly successful on semidetached class, while it tends to predict overcontact binaries as semidetached.

Training of the ML model is one of the important steps. The Kepler and TESS light curves that we used for training should be modeled in the same way as the Gaia light curves. This is the reason why the model was successful in classifying the test data, but  partially failed in classifying the Gaia data. We plan to explore additional modeling techniques, including smoothing and fitting analytic models, for improved consistency and interpretability in future studies.

Although model accuracy could probably be improved with a larger and more homogeneous dataset, our primary focus was to establish a proof of concept and identify potential challenges to this classification task. As part of our ongoing work, additional observation archives are being explored to expand the training dataset and improve the robustness and performance of the model.

We used "phase folding method" for the acquisition of the relevant light curve information. Harmonic analysis can be used to determine characteristics of the light curves. Integration of spot-induced effects as well as modulations caused by reflection and ellipsoidal deformation in close binary systems could enhance the precision of future models, particularly for more detailed analysis of system parameters. Spot-induced variations are generally low-amplitude compared to the primary eclipse features in eclipsing binaries. Despite such variations, classification accuracy is found to be similar across diverse types of light curves with varying levels of spot activity.


\begin{thebibliography}{7}
\expandafter\ifx\csname natexlab\endcsname\relax\def\natexlab#1{#1}\fi

\bibitem[{B\'{o}di \& Hajdu(2021)}]{Bodi_2021}
B\'{o}di, A. \& Hajdu, T., Classification of OGLE Eclipsing Binary Stars Based on Their Morphology Type with Locally Linear Embedding. 2021, {\it The Astrophysical Journal Supplement Series}, {\bf 255}, 1, DOI: 10.3847/1538-4365/ac082c

\bibitem[{{\v{C}okina} {et~al.}(2021){\v{C}okina}, Maslej-Kre\v{s}\v{n}\'{a}kov\'{a}, Butka, \& Parimucha}]{Cokina_2021}
{\v{C}okina}, M., Maslej-Kre\v{s}\v{n}\'{a}kov\'{a}, V., Butka, P., \& Parimucha, v., Automatic classification of eclipsing binary stars using deep learning methods. 2021, {\it Astronomy and Computing}, {\bf 36}, 100488, DOI: 10.1016/j.ascom.2021.100488

\bibitem[{Daza-Perilla {et~al.}(2023)Daza-Perilla, Gramajo, Lares, Palma, Ferreira~Lopes, Minniti, \& Clari\'{a}}]{Daza_2023}
Daza-Perilla, I.~V., Gramajo, L.~V., Lares, M., {et~al.}, {Automated classification of eclipsing binary systems in the VVV Survey}. 2023, {\it Monthly Notices of the Royal Astronomical Society}, {\bf 520}, 828, DOI: 10.1093/mnras/stad141

\bibitem[{{Kochoska, A.} {et~al.}(2017){Kochoska, A.}, {Mowlavi, N.}, {Pr\v{s}a, A.}, {Lecoeur-Ta\"{\i}bi, I.}, {Holl, B.}, {Rimoldini, L.}, {S\"{u}veges, M.}, \& {Eyer, L.}}]{Kochoska_2017}
{Kochoska, A.}, {Mowlavi, N.}, {Pr\v{s}a, A.}, {et~al.}, Gaia eclipsing binary and multiple systems. A study of detectability and classification of eclipsing binaries with Gaia. 2017, {\it A\&A}, {\bf 602}, A110, DOI: 10.1051/0004-6361/201629957

\bibitem[{Matijevi\v{c} {et~al.}(2012)Matijevi\v{c}, Pr\v{s}a, Orosz, Welsh, Bloemen, \& Barclay}]{Matijevic_2012}
Matijevi\v{c}, G., Pr\v{s}a, A., Orosz, J.~A., {et~al.}, Kepler Eclipsing Binary Stars. III. Classification of Kepler Eclipsing Binary Light Curves with Locally Linear Embedding. 2012, {\it The Astronomical Journal}, {\bf 143}, 123, DOI: 10.1088/0004-6256/143/5/123

\bibitem[{{Mowlavi, N.} {et~al.}(2023){Mowlavi, N.}, {Holl, B.}, {Lecoeur-Ta\"{\i}bi, I.}, {Barblan, F.}, {Kochoska, A.}, {Pr\v{s}a, A.}, {Mazeh, T.}, {Rimoldini, L.}, {Gavras, P.}, {Audard, M.}, {Jevardat de Fombelle, G.}, {Nienartowicz, K.}, {Garc\'{\i}a-Lario, P.}, \& {Eyer, L.}}]{Mowlavi_2023}
{Mowlavi, N.}, {Holl, B.}, {Lecoeur-Ta\"{\i}bi, I.}, {et~al.}, Gaia Data Release 3 - The first Gaia catalogue of eclipsing-binary candidates. 2023, {\it A\&A}, {\bf 674}, A16, DOI: 10.1051/0004-6361/202245330

\bibitem[{{S\"{u}veges, M.} {et~al.}(2017){S\"{u}veges, M.}, {Barblan, F.}, {Lecoeur-Ta\"{\i}bi, I.}, {Pr\v{s}a, A.}, {Holl, B.}, {Eyer, L.}, {Kochoska, A.}, {Mowlavi, N.}, \& {Rimoldini, L.}}]{Suveges_2017}
{S\"{u}veges, M.}, {Barblan, F.}, {Lecoeur-Ta\"{\i}bi, I.}, {et~al.}, Gaia eclipsing binary and multiple systems. Supervised classification and self-organizing maps. 2017, {\it A\&A}, {\bf 603}, A117, DOI: 10.1051/0004-6361/201629710

\end{thebibliography}

\end{document}